\documentclass[11pt,a4paper]{article}
\pdfoutput=1   
\usepackage{amsmath,amsfonts,amssymb,amsbsy}
\usepackage{mathrsfs,latexsym}
\usepackage{graphicx}
\usepackage{xcolor}
\usepackage{booktabs}
\usepackage{multirow}
\usepackage{multicol}
\usepackage{subfig}
\usepackage{placeins}
\usepackage{bm}
\usepackage{siunitx}
\usepackage[T1]{fontenc}%
\usepackage[utf8]{inputenc}%
\usepackage{lmodern}%
\usepackage{textcomp}%
\usepackage{longtable}%
\usepackage{alpha}
\usepackage{macros_alpha}
\renewcommand{\strange}{\mathrm{strange}}
\usebiblio{latticen}

\usepackage{tikz}
\usepackage{tikzscale}
\usetikzlibrary{calc}
\usetikzlibrary{plotmarks}
\usetikzlibrary{decorations.markings}
\usetikzlibrary{fit}

\newcommand{\dvec}{\bm{d}}

\newcommand{\cm}{{\mathrm {cm}}}

\begin{document}
\newcommand*{\BUILDFIGS}{}%

\preprintno{%
CP3-Origins-2018-5 DNRF90\\
HIM-2018-01\\
MITP/18-010}

\title{Determination of $s$- and $p$-wave $I=1/2$ $K\pi$ scattering amplitudes in $N_{\mathrm{f}}=2+1$ 
	lattice QCD}

\author[cmu]{Ruair\'i~Brett}
\author[usden]{John~Bulava}
\author[kentucky]{Jacob~Fallica}
\author[him]{Andrew~Hanlon}
\author[mainz]{Ben~H\"{o}rz}
\author[cmu]{Colin~Morningstar}

\address[cmu]{Department~of~Physics, Carnegie~Mellon~University, 
              Pittsburgh, PA~15213, USA}
\address[usden]{Dept.~of Mathematics~and~Computer~Science and CP3-Origins, 
                University of Southern Denmark, Campusvej 55, 5230 Odense M, Denmark}
\address[kentucky]{Department of Physics and Astronomy, University of Kentucky,
              Lexington, KY 40506, USA}
\address[him]{Helmholtz-Institut Mainz,
                Johannes Gutenberg-Universit\"at, 55099 Mainz, Germany}
\address[mainz]{PRISMA Cluster of Excellence and Institute for Nuclear Physics, 
                Johannes Gutenberg-Universit\"at, 55099 Mainz, Germany}

\begin{abstract}

	The elastic $I=1/2$, $s$- and $p$-wave kaon-pion scattering amplitudes are 
	calculated using a single ensemble of anisotropic lattice QCD 
	gauge field configurations with $N_{\mathrm{f}} = 2+1$ flavors 
	of dynamical Wilson-clover fermions at $m_{\pi} = 230\mathrm{MeV}$. A large 
	spatial extent of $L = 3.7\mathrm{fm}$ enables a good energy resolution while
	partial wave mixing due to the reduced symmetries of the finite volume
	is treated explicitly. 
	The $p$-wave amplitude is well described by a Breit-Wigner 
	shape with parameters 
	$m_{K^{*}}/m_{\pi} = 3.808(18)$ and $g^{\mathrm{BW}}_{K^{*}K\pi} = 5.33(20)$ which 
	are insensitive to the inclusion of $d$-wave mixing and variation 
	of the $s$-wave parametrization. 
	An effective range description of the near-threshold $s$-wave 
	amplitude yields $m_{\pi}a_0 = -0.353(25)$.  

\end{abstract}

\begin{keyword}
lattice QCD, meson scattering
	\PACS{%
12.38.Gc\sep 
11.15.Ha\sep 
11.30.Rd\sep 
13.30.Eg\sep 
13.75.Lb\sep 
13.85.Dz\sep 
14.40.Be 
}    
\end{keyword}

\maketitle

\section{Introduction}

Elastic $K\pi$ scattering amplitudes are essential to several 
current frontiers in the phenomenology of the Standard Model of particle 
physics. 
For example, precision tests of lepton universality performed at CERN by the LHCb 
collaboration using decays in which the elastic 
$I(J^P)=\frac{1}{2}(1^{-})$ 
$K^*(892)$ resonance is produced exhibit deviations between theory and 
experiment in 
$R_{K^*} = \mathrm{BR}(B\rightarrow K^{*}\mu^{+}\mu^{-})/\mathrm{BR}(B\rightarrow K^{*} \mathrm{e}^{+}\mathrm{e}^{-})$ at the $(2.1-2.5)\sigma$ 
level~\cite{Aaij:2017vbb}. Although the hadronic form 
factors involved in these branching fractions cancel in the ratio, precise 
lattice QCD predictions are desirable. Existing lattice calculations of these 
 form factors however do not treat the
$K^{*}$ as a unstable particle~\cite{Horgan:2013pva,Horgan:2013hoa}. The 
theoretical formalism to extract form factors correctly treating the unstable 
nature of the $K^{*}$ is well known~\cite{Agadjanov:2016fbd,Lellouch:2000pv} 
and requires the elastic $p$-wave $K\pi$ scattering amplitude calculated in 
this work. 

In addition to the $K^{*}(892)$ resonance, the nature and existence of the 
low-lying broad $s$-wave $K_0^{*}(800)$ resonance is not 
clear~\cite{Patrignani:2016xqp}. The  
amplitudes calculated in this work may provide information on the 
quark-mass dependence of these resonance poles and confront expectations 
from chiral effective theories~\cite{Doring:2012eu,Danilkin:2011fz,
Doring:2011nd,Nebreda:2010wv,Bernard:1990kx,Bernard:1990kw}.  

Finally, apart from study of the resonances, the $K\pi$ $s$-wave scattering lengths are of phenomenological interest as a precision Standard Model test. The DIRAC 
experiment at CERN has produced promising results for these quantities 
using $\pi K$ `atoms' and plans to achieve 5\% accuracy~\cite{DIRAC:2016rpv}. The $I=1/2$, $s$-wave scattering length calculated in this work is therefore 
an important step toward an accurate and precise determination of these 
scattering lengths using lattice QCD.

While lattice QCD is a proven tool to determine hadronic 
properties from first principles, real-time hadron-hadron scattering amplitudes 
are significantly more difficult to calculate than single-hadron properties 
due to the Euclidean space-time lattice~\cite{Maiani:1990ca}. However, a 
particularly successful approach developed by L\"{u}scher circumvents 
this difficulty by inferring elastic scattering amplitudes from the deviation 
of finite-volume hadron-hadron energies from their non-interacting 
values~\cite{Luscher:1990ux}. This method has been extended
to moving frames~\cite{Rummukainen:1995vs,Kim:2005gf}, 
particles with spin~\cite{Gockeler:2012yj,Morningstar:2017spu,Briceno:2013lba,
Briceno:2014oea}, and coupled 
two-hadron channels~\cite{He:2005ey}. 
Progress toward the full extension to three-hadron
amplitudes has been made in Refs.~\cite{Doring:2018xxx,Mai:2017bge,Hammer:2017kms,
Hammer:2017uqm,Hansen:2015zga,Hansen:2014eka, Hansen:2016ync,Hansen:2016fzj,
Briceno:2017tce,Polejaeva:2012ut,Briceno:2012rv} and 
amplitudes with an external current can also be 
calculated~\cite{Briceno:2015tza,Briceno:2015csa,Briceno:2012yi,Hansen:2012tf,
Meyer:2011um,Lellouch:2000pv}.  
Alternative approaches 
to handle inclusive decays which do not employ the finite volume 
have been proposed recently in 
Refs.~\cite{Hansen:2017mnd,Hashimoto:2017wqo}. 

In addition to this theoretical progress, algorithmic 
advances~\cite{Peardon:2009gh,Morningstar:2011ka} and Moore's 
law 
have resulted in 
considerable  
progress in lattice QCD calculations of finite-volume hadron-hadron energy 
spectra, and thus by extension scattering amplitudes as well. 
The state of such calculations has been 
reviewed recently in Ref.~\cite{Briceno:2017max}. Lattice determinations of 
elastic meson-meson amplitudes are 
increasingly precise~\cite{Alexandrou:2017mpi,Bali:2015gji,Fu:2016itp,Feng:2014gba,Feng:2010es,
Orginos:2015aya,Beane:2011sc,Pelissier:2012pi,Aoki:2011yj,
Dudek:2012xn,Dudek:2012gj,Lang:2016jpk,Lang:2015hza,
Lang:2014yfa,Mohler:2013rwa,Prelovsek:2013ela,Mohler:2012na,Lang:2012sv,
Lang:2011mn,Guo:2016zos,Helmes:2017smr,Liu:2016cba,Helmes:2015gla,
Bulava:2016mks} while those of meson-baryon and baryon-baryon  systems
have considerably larger errors~\cite{Fukugita:1994ve, Meng:2003gm, Torok:2009dg, 
Detmold:2015qwf,Lang:2012db,Lang:2016hnn}. First results 
with coupled meson-meson channels have also been  
performed~\cite{Moir:2016srx,Briceno:2016mjc,Briceno:2016kkp,Dudek:2016cru,Wilson:2015dqa,
Wilson:2014cna}. 

The $K\pi$ amplitudes described in this work present 
additional difficulties 
compared to the $\pi\pi$ case. 
Due to the reduced symmetry (compared to infinite volume) of the finite toroidal volume  in which 
our simulations are performed, partial wave amplitudes with 
different orbital angular momenta contribute to the energy shift of a single 
finite-volume energy. 
The pattern of this partial wave
mixing is more complicated 
at non-zero total momentum if the hadrons are not identical. A practical theoretical and statistical treatment of these effects has been 
proposed recently in Ref.~\cite{Morningstar:2017spu}, which details the 
procedure we follow here.

The main results of this work are parametrizations of the $s$- and $p$-wave 
$I=1/2$ elastic $K\pi$ scattering amplitudes. The $p$-wave amplitude 
is well-described by a Breit-Wigner, as expected in the presence of a 
narrow $K^*(892)$ resonance while the energy dependence of the $s$-wave 
amplitude can be fit with several ansatze including a Breit-Wigner. All 
parameters from these fits are listed in Tab.~\ref{t:fit} and 
plots of several of them are shown in Fig.~\ref{f:cot}.  
In addition to these parametrizations, we provide in Tab.~\ref{t:res} the 
finite-volume energies 
and box matrix elements (which are defined in Eq.~\ref{e:box}) to enable 
additional future parametrizations. 

The remainder of this paper is organized as follows. In Sec.~\ref{s:meth} we 
outline the ensemble of gauge 
field configurations, methods for calculating finite-volume two-hadron energies,
and the relation of those energies to infinite volume scattering amplitudes. 
This is followed by Sec.~\ref{s:res}, where results are presented, and Sec.~\ref{s:concl} which contains conclusions. 

\section{Methods}\label{s:meth}

We employ the single ensemble of anisotropic $\nf = 2+1$ Wilson clover fermions
 used previously in Ref.~\cite{Bulava:2016mks} for 
 elastic pion-pion scattering. Much 
of the procedure developed there to determine the finite-volume two-hadron energies is taken over in this work. 
However, the relation of finite-volume energies to the desired 
amplitudes is complicated significantly 
with respect to Ref.~\cite{Bulava:2016mks} so the analysis methods proposed  
in Ref.~\cite{Morningstar:2017spu} must be employed.  

\subsection{Ensemble Details}\label{s:ens}

The ensemble of anisotropic gauge configurations used in this work is detailed 
in Refs.~\cite{Lin:2008pr,Edwards:2008ja} but we review the salient points here. The gauge action is Symanzik and tadpole improved at tree level and a clover 
term is added to the Wilson action for fermions. The spatial gauge links appearing in the fermion action are stout smeared.
Properties of the ensemble relevant for this work are listed in 
Tab.~\ref{t:ens}. While all our dimensionful results are expressed as dimensionless ratios using $m_{\pi}$, an indicative value of the 
lattice scale is obtained (as in Ref.~\cite{Bulava:2016mks}) by demanding that 
the kaon mass take its physical value. Such a (mass-dependent) scale setting 
gives $a_t = 0.033357(59)\mathrm{fm}$.
\begin{table}
\centering
\begin{tabular} {c c c c c c}
  \toprule
$(L/a_s)^3 \times (T/a_t)$ & $N_{\rm cfgs}$ & $a_t m_\pi$ & $a_t m_K$ & $a_t m_\eta$ & $\xi$ \\ \midrule
$32^3 \times 256$ & $412$ & $0.03938(19)$ & $0.08354(15)$ & $0.1010(37)$ & $3.451(11)$ \\ \bottomrule
\end{tabular}
	\caption{\label{t:ens}Details of the ensemble used in this work. Pion and kaon masses are taken from Ref.~\cite{Bulava:2016mks} while the
	determination of $m_{\eta}$ is discussed in the text. The renormalized
	anisotropy $\xi$ is set using the pion dispersion relation and also taken
	from Ref.~\cite{Bulava:2016mks}. Setting the scale
	with the kaon mass gives $a_t = 0.033357(59)\mathrm{fm}$.}
\end{table}

This work is based on $N_{\mathrm{cfg}}=412$ configurations which are separated 
by $20$ Hybrid Monte Carlo (HMC) trajectories of length $\tau =1$. While no 
statistically significant autocorrelations are observed in any of the 
correlation functions we consider here, in order to mitigate the effect of 
autocorrelation on our estimates of the statistical error we bin the data in bins of size $N_{\mathrm{bin}}=2$.
Using this binned data set, statistical errors are estimated using the 
bootstrap procedure with $N_{B} = 800$ bootstrap samples. 

\begin{figure}
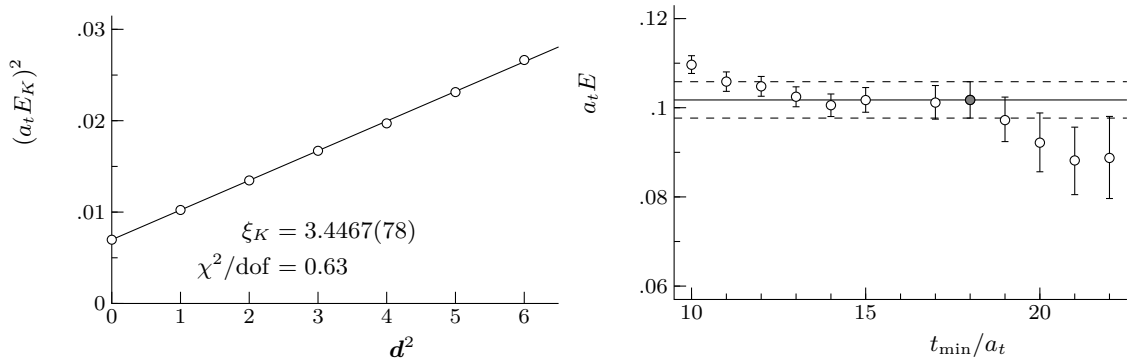

  \centering
	\includegraphics[width=0.49\textwidth]{figures/kaon_dispRel.tikz}
	\includegraphics[width=0.49\textwidth]{figures/eta_pSq0_tmin.tikz}
	\caption{\label{f:keta} \textbf{Left}: linear fit to the energies of 
	single-kaon correlation functions to determine the kaon anisotropy $\xi_K$, 
	which is consistent with $\xi_{\pi}$ from the pion dispersion relation. 
	\textbf{Right}: $\tmin$-plot for the determination of $m_{\eta}$ with $\tmax = 26a_t$. The chosen fit range is indicated with a filled circle. This analysis involves a GEVP and 
	is discussed further in the text.}
\end{figure}

Determination of the pion and kaon masses is discussed in 
Ref.~\cite{Bulava:2016mks} and we take over those values here. We additionally 
employ the renormalized anisotropy ($\xi=a_s/a_t=\xi_{\pi}$) determined in 
Ref.~\cite{Bulava:2016mks} by enforcing the correct relativistic dispersion 
relation for the pion. As discussed in 
Ref.~\cite{Wilson:2015dqa}, $\xi$ is insensitive to the hadron whose 
dispersion relation is used. For example, our determination of $\xi_K$ using
the kaon dispersion relation is shown in Fig.~\ref{f:keta} and agrees well with
$\xi_{\pi}$. The linear fit used to determine $\xi_K$ is shown
in Fig.~\ref{f:keta}, in which the individual energies are obtained from 
single-exponential fits to the relevant correlation functions which ignore the 
finite temporal extent $T$. As demonstrated in 
Ref.~\cite{Bulava:2016mks}, at the current level of statistical precision, we 
find that  finite-$T$ effects are negligible. These single exponential fits are 
performed over a range $[\tmin,\tmax]$, so that the level of unwanted 
excited state contamination can be monitored by varying $\tmin$. If the 
fitted energies do not exhibit statistically significant variation for a range of $\tmin$, 
systematic errors due to excited states are smaller than the statistical errors. The $\tmin$ plots illustrating these plateaux for all single-$K$ levels 
used in determining $\xi_{K}$ are shown in Fig.~\ref{f:xik} of App.~\ref{a:k}. 
We also check that finite-$T$ effects are small by observing that 
energies obtained from fit form $A\exp (-E t)$ are indistinguishable from 
those obtained using fit form $A [ \exp(-Et) + \exp(-E(T-t))]$ plus other 
terms that can occur in two-meson correlators.

In addition to the pion and kaon masses, we require an estimate of $m_{\eta}$.   These three hadron masses determine the position of relevant inelastic 
thresholds for $I=1/2$, $S=1$ kaon-pion scattering, which is the focus of this work. 
To determine $m_{\eta}$, we solve a generalized eigenvalue problem (GEVP) 
which includes two single-site interpolating operators with flavor content $\bar{u}u + \bar{d}d$ and $\bar{s}s$. The GEVP 
\begin{align}\label{e:gevp}
	C(t_d) \, v(t_0,t_d) = \lambda(t_0, t_d) C(t_0) \, v(t_0,t_d) 
\end{align}
is solved once for a single choice of the diagonalization times 
$(t_0,t_d)=(9a_t, 18a_t)$, where $C(t)$ is the $2\times 2$ correlation matrix composed of the light and strange interpolators. The eigenvector corresponding to the largest eigenvalue is used to 
rotate the correlation matrix and obtain a single diagonal correlation function 
which has optimal overlap onto the 
ground state~\cite{Michael:1982gb}. Single exponential fits to this optimized 
ground state correlation function for varying $\tmin$ are displayed in 
Fig.~\ref{f:keta}. These fitted energies vary little with $(t_0,t_d)$ and are 
insensitive to an enlargement of the GEVP operator basis. As a zero-strangeness 
isoscalar, $\eta$-meson correlation functions 
contain fully disconnected quark lines, our estimation of which is discussed in 
Sec.~\ref{s:corr}. These relative quark lines start and end at the same time
 and are estimated using non-maximal time dilution, in which each dilution 
 projector has support on every sixteenth timeslice. Since we only employ a 
 single combination of stochastic sources, our estimate of $\eta$-meson correlation functions at a separation of precisely $t=16a_t$ is poorly estimated compared to the other points. Time separation $t=16a_t$ is therefore removed from all fits to 
 $\eta$-meson correlators.  

Using the masses $m_{\pi}$, $m_{K}$, and $m_{\eta}$ the 
relevant inelastic thresholds 
are given in Tab.~\ref{t:thr}. While the formalism discussed in Sec.~\ref{s:amp} can relate energies above 
inelastic two-hadron thresholds to the corresponding coupled-channel scattering 
amplitude, the situation above three-hadron thresholds is more complicated~\cite{Polejaeva:2012ut,Hansen:2014eka}.
On this ensemble $\pi\pi K$ is 
the lowest inelastic threshold and sits below $\eta K$, so we are able to 
treat elastic $\pi K$ scattering only. 
A convenient parameter delineating the elastic region is $\tilde{E} = 
(E_{\cm} - m_K)/m_{\pi}$. As indicated in the table, the elastic region of 
interest extends over the range $ 1 < \tilde{E} < 2$. 
\begin{table}
\centering
	\begin{tabular}{c c c c}
	  \toprule
		& $a_t E_{\rm th}$ & $E_{\rm th} / m_\pi$ & $(E_{\rm th} - m_K)/m_{\pi}$ \\
		\midrule
		$\pi K$ & $0.12293(24)$ & $3.121(11)$ & $1$ \\
		$\pi \pi K$ & $0.16233(40)$ & $4.121(11)$ & $2$ \\
		$\eta K$ & $0.1845(37)$ & $4.664(99)$ & $2.553(96)$ \\
		\bottomrule
\end{tabular}
\caption{\label{t:thr}Relevant inelastic thresholds for $I=\frac{1}{2}$, $S=1$ 
	kaon-pion scattering. Since the lowest inelastic threshold contains three 
	hadrons, we treat elastic scattering only.}
\end{table}

\subsection{Correlation function construction}\label{s:corr} 

As discussed in Sec.~\ref{s:amp}, infinite volume elastic scattering 
amplitudes are related to finite-volume two-hadron energies. These energies 
are determined in lattice QCD simulations by fitting the exponential fall-off 
of temporal correlation functions between suitable interpolating operators.  
In order to employ two-hadron interpolating operators in which each hadron 
has definite momentum, and to treat Wick contractions where some quark lines 
start and end at the same time, all-to-all quark propagators between each 
spacetime point are required. 

Such all-to-all propagators are intractable to evaluate directly but efficient
stochastic estimators can be constructed for  
propagators projected onto the space spanned by the $N_{\rm ev}$ 
lowest eigenmodes 
of the (stout-smeared) gauge-invariant three-dimensional Laplace operator~\cite{Peardon:2009gh,Morningstar:2011ka}.
We refer to this as the LapH subspace and  
this projection is simply a particular form of quark smearing, which 
has long been used to reduce the amount of unwanted excited state overlap 
in hadronic interpolating operators. 

These stochastic estimators introduce noise into the LapH subspace and may 
be improved via dilution~\cite{Foley:2005ac}, in which a set of complete orthogonal projectors 
is specified in time, spin, and Laplacian eigenvector indices. 
We differentiate quark lines which start and end at the same time (relative 
lines) from those which start and end at different times (fixed lines), and it is 
beneficial to adopt different dilution schemes for fixed and relative lines.  
For this work we employ the same quark smearing ($N_{\rm ev} = 264$) and 
dilution schemes as Ref.~\cite{Bulava:2016mks}. In addition to the light quark 
inversions performed there, we require a single independent fixed 
strange 
line. All correlators are estimated using a minimal number of stochastic 
sources, and only a single permutation of these sources is employed.  
Although additional Dirac matrix inversions are performed in order to 
construct correlators for other systems, the results of this work employ 
three fixed light quark lines, one fixed strange quark line, and a single 
relative light quark line. Given the dilution schemes employed here, this 
work therefore requires $N_{\rm light} = 1280$ light Dirac matrix inversions 
and $N_{\rm strange} = 256$ strange inversions on each gauge configuration. 
The determination of $m_{\eta}$ discussed in Sec.~\ref{s:ens} additionally uses a single relative strange line requiring another 
$N_{\rm strange} = 512$ strange inversions.  

Using the source and sink functions defined in Ref.~\cite{Morningstar:2011ka}, all Wick 
contractions (also enumerated in Ref.~\cite{Morningstar:2011ka}) for correlation
functions between single-meson and meson-meson interpolators may be 
efficiently evaluated. The single-meson and meson-meson 
operators employed here are taken from Ref.~\cite{Morningstar:2013bda} 
and transform irreducibly according to the appropriate finite-volume 
symmetry group. We consider $\pi K$ operators at zero total momentum 
$\boldsymbol{d}^2 = (L/2\pi)^2 \boldsymbol{P}_{\mathrm{tot}}^2 $ as well as all non-zero 
on-axis, planar-diagonal, and cubic diagonal total momenta up to 
$\boldsymbol{d}^2 \le 4$. 

Our interpolating operators therefore transform irreducibly according to the 
appropriate little group for each total momentum ray. Due to the reduced 
symmetry of the finite periodic spatial volume, a single infinite-volume 
irrep (labeled by orbital angular momentum $\ell$) will be subduced onto 
possibly several finite volume irreps, which are denoted by $\Lambda$. 
The $\ell$th partial wave may also occur multiple times in a particular 
finite volume irrep.
This subduction pattern is illustrated in Tab.~\ref{t:irr} for the irreps 
considered in this work. 
\begin{table}
\centering
\begin{tabular}{c c l}
  \toprule
	$\dvec$ & \textbf{$\Lambda$} & $\ell$ \\
	\midrule
	$(0,0,0)$ &$A_{1g}$& 0, 4, \ldots \\
&$T_{1u}$& 1, 3, \ldots \\
\midrule%
	$(0,0,n)$ &$A_1$& 0, 1, 2, \ldots \\
&$E$& 1, 2, 3, \ldots \\
\midrule%
	$(0,n,n)$ &$A_1$& 0, 1, 2, \ldots \\
&$B_1$& 1, 2, 3, \ldots \\
&$B_2$& 1, 2, 3, \ldots \\
\midrule%
	$(n,n,n)$ &$A_1$& 0, 1, 2, \ldots \\
&$E$& 1, 2, 3, \ldots \\
\bottomrule
\end{tabular}
\caption{\label{t:irr} Irreps $\Lambda$ of the appropriate little group for various 
	total momenta $\boldsymbol{P}_{\rm tot} = (2\pi/L)\dvec$ (where $\dvec$ is a vector of integers) considered in this work. We consider $K\pi$
	systems at rest as well as those with non-zero total on-axis, planar-diagonal, and cubic-diagonal momenta. These momentum classes are listed in the first column, where $n\in \mathbb{Z}$ is an arbitrary integer.}
\end{table}
Increased complication with respect to the pion-pion case due to non-identical
particles is now evident. While even
and odd partial waves do not contribute to the same irrep at zero total 
momentum, there are no non-zero momentum irreps to which $\ell=0$ contributes but not $\ell=1$. 

In each of the irreps listed in Tab.~\ref{t:irr}, we form a temporal 
correlation matrix from which the finite-volume spectrum is extracted. Assuming 
the presence of a narrow $K^*(892)$ resonance and allowing for the possibility 
of an additional $s$-wave resonance, these correlation matrices are composed 
of (non-displaced) single-hadron operators as well as kaon-pion operators with
various individual momenta. Using estimates based on $m_K$, $m_\pi$ and $L$, 
roughly $2-6$ irreducible $K\pi$ 
interpolating operators corresponding to the lowest non-interacting states are included in 
each irrep, together with $1-2$ single-hadron operators. These operators are 
intended to have large overlap onto all elastic states of interest, as well as 
a few states above inelastic threshold. Using a larger basis than strictly necessary enables a check of the stability of the spectrum as a few higher-lying operators are removed.  
Full specification of the operators included in each irrep is given in 
App.~\ref{a:ops}. 

\subsection{Finite-Volume Spectrum Determination}\label{s:nrg}
\begin{figure}
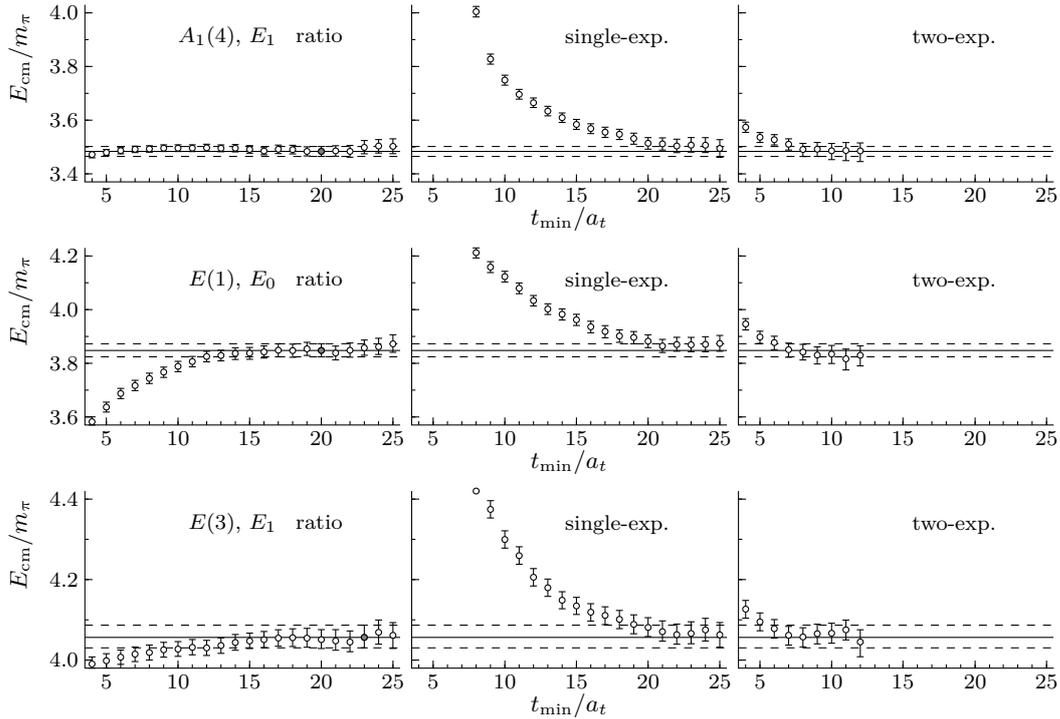

	\ifdefined\BUILDFIGS
	\includegraphics{figures/pSq4-A11_fitSyst.tikz}
	\includegraphics{figures/pSq1-E0_fitSyst.tikz}
	\includegraphics{figures/pSq3-E1_fitSyst.tikz}
	\fi
	\caption{\label{f:fit1} Comparison of ratio, single exponential, and two-exponential fits
	for a selection of levels throughout the elastic region. Each row corresponds
	to the three fits for a single level specified in the left column as
	`$\Lambda(\boldsymbol{d}^2), \, E_n$', denoting the $n$th level in finite volume
	irrep $\Lambda$ with total momentum $\boldsymbol{d}^2$. Each plot 
	shows the variation of the fitted energy with $\tmin$, the lower end of 
	the fitting range, with the chosen fit indicated with a filled symbol.}
\end{figure}

Given the correlation matrices discussed in Sec.~\ref{s:corr}, we turn now 
to methods for extracting finite-volume spectra from them. 
As discussed in Sec.~\ref{s:ens} in the determination of single hadron masses, 
we can safely neglect finite temporal extent effects. Since several excited 
states are desired in addition to the ground state in each irrep, 
GEVP methods are employed which solve Eq.~\ref{e:gevp} once for a single 
choice of $(t_0,t_d)$ and a correlation matrix of size $N_{\mathrm{op}}$. 
The operators included in the GEVP are given in Tab.~\ref{t:ops} in 
App.~\ref{a:ops}. Any variation of the spectrum with $(t_0,t_d)$ and 
$N_{\mathrm{op}}$ implies a systematic uncertainty whose magnitude 
must be assessed.

Using the GEVP eigenvectors $\{v_n\}$ the correlation matrix is rotated 
\begin{align}
	\hat{C}_n(t) = (v_n, C(t)v_n),
\end{align}
where the outer parentheses on the RHS denote an inner product over the GEVP 
index. $\hat{C}_n(t)$ is a diagonal correlation function with optimal 
overlap onto finite-volume energy level $n$~\cite{Michael:1982gb}. As discussed in Sec.~\ref{s:amp}, 
the signal of interest is the deviation of the finite-volume two-hadron energies
from their non-interacting counterparts. To this end, the energy difference 
$\Delta E$ is extracted directly by constructing the ratio 
\begin{align}
	R_n(t) = \frac{\hat{C}_n(t)}{C_\pi(\boldsymbol{d}_{\pi}^2, t) \, 
	C_{K}(\boldsymbol{d}_{K}^2,t)}
\end{align}
where the nearest non-interacting state to energy level $n$ consists of a pion with momentum $\boldsymbol{d}_{\pi}^2$ and a kaon with momentum 
$\boldsymbol{d}_{K}^2$. Single-exponential correlated-$\chi^2$ fits are 
performed to the ansatz $R_n(t) = A_n \mathrm{e}^{-\Delta E_n t}$. 

For weakly interacting levels where $\Delta E_n$ is small, these ratio fits 
generally have considerably smaller excited contamination than fits to 
$\hat{C}_n(t)$ directly. However, the identification of plateau with the ratio 
fits is complicated somewhat because the contributions from unwanted 
higher-lying states do not necessarily enter with a positive sign, as they do 
in $\hat{C}_n(t)$~\cite{Bulava:2016mks}. These `bumps' are evident in $\tmin$-plots for levels which exhibit significant deviation from the non-interacting 
energies. Nonetheless, taking these bumps into account results in consistent 
energies and statistical errors for these levels between ratio fits and exponential fits to $\hat{C}_n(t)$. 
Ratio fits are compared to fits to $\hat{C}_n(t)$ directly using single-exponential and two-exponential ansatze in Fig.~\ref{f:fit1}.

\begin{figure}
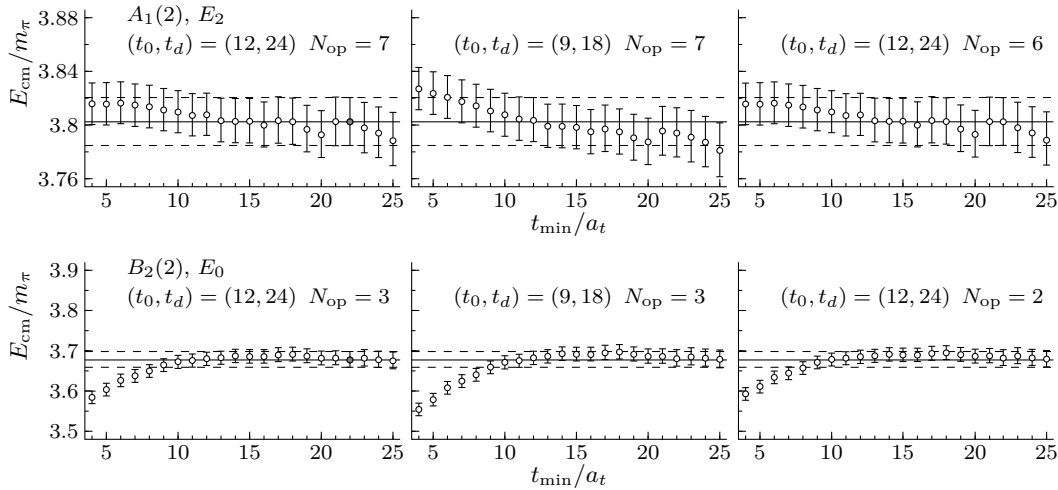

	\ifdefined\BUILDFIGS
	\includegraphics{figures/pSq2-A12_gevpSyst.tikz}
	\includegraphics{figures/pSq2-B20_gevpSyst.tikz}
	\fi
	\caption{\label{f:fit2}Comparison of ratio fits for different 
	GEVP parameters for a selection of levels. As in Fig.~\ref{f:fit1}, each row 
	corresponds to a different energy, denoted in the left column. The GEVP 
	bases are given in Tab.~\ref{t:ops} where bases with one fewer operator are 
	formed by discarding the last entry in each operator list. Each plot 
	shows the variation of the fitted energy with $\tmin$, the lower end of 
	the fitting range, while the chosen fit is indicated with a filled circle. GEVP systematics for all other levels are less pronounced than those shown here.}   
\end{figure}

As in Ref.~\cite{Bulava:2016mks} ratio fits are employed for the final 
amplitude analysis. Similar to the  
single-hadron operator fits discussed in Sec.~\ref{s:ens}, $\tmin$ must be chosen so that systematic errors due
to unwanted excited state contamination are smaller than statistical ones. Our general critera for choosing a suitable $\tmin$ are 
$\chi^2/{\rm d.o.f.} < 1.7$ and 
\begin{align*}
	\Delta E_{\rm fit} (\tmin) - \Delta E_{\rm fit} (\tmin - \delta_t) < 
	\sigma(\tmin), 
\end{align*}
where $\Delta E_{\rm fit}(\tmin)$ is the energy difference obtained from the 
fit range $\left[\tmin, \tmax\right]$, $\sigma(\tmin)$ its statistical error, and 
$\delta_t = 4a_t$. 
For 
these fits we additionally require that any variation with $(t_0,t_d)$ or 
$N_{\mathrm{op}}$ is also smaller than the statistical error. Generally the variation of the energies 
with the GEVP parameters is small, as illustrated in Fig.~\ref{f:fit2}.

\subsection{Amplitudes from finite-volume energies}\label{s:amp}

After determining $\Delta E_n$ as described in Sec.~\ref{s:nrg}, we reconstruct
the finite volume energies via 
\begin{align}\label{e:erf}
	a_tE_n = a_t \Delta E_n + \sqrt{a_t^2m_\pi^2 + \left(\frac{2\pi a_s}{\xi L}\right)^2 \boldsymbol{d}_{\pi}^2} +  \sqrt{a_t^2m_K^2 + \left(\frac{2\pi a_s}{\xi L}\right)^2 \boldsymbol{d}_{K}^2}.
\end{align}
These energies are determined in the `lab' frame in which the $K\pi$ system 
may have non-zero total momentum. These lab-frame energies are related to 
quantities in the center-of-mass frame by
\begin{align}\label{e:kin}
	E_{\cm} = \sqrt{E^2 - \boldsymbol{P}_{\mathrm{tot}}^2}, \quad 
	q^2_{\cm} = \frac{1}{4}E_{\cm}^2 - \frac{1}{2}(m_{\pi}^2 + m_{K}^2) + \frac{(m_\pi^2 - m_{K}^2)^2}{4E^2_{\cm}}, 
\end{align}
where $E$ is the lab frame energy. 

The relation between two-particle center-of-mass energies and the 
infinite-volume elastic
scattering amplitude may be expressed as~\cite{Morningstar:2017spu} 
\begin{align}\label{e:box}
	\mathrm{det} [\tilde{K}^{-1}(E_{\cm}) - B^{(\Lambda, \boldsymbol{d})}(E_{\cm})] = 0 
\end{align}
which holds up to corrections which are exponentially suppressed in the spatial 
extent $L$. For the elastic scattering of two spinless particles, 
$\tilde{K}^{-1}$ and $B$ are infinite-dimensional matrices in both $\ell$ and 
$n_{\mathrm{occ}}$, an index enumerating the possibly multiple occurrences of 
a single partial wave in a particular irrep. Note that $B$ depends on the  
total momentum class and irrep. Expressions and numerical programs for evaluation of the $B$-matrix elements are provided in Ref.~\cite{Morningstar:2017spu}. 

For a unitary elastic scattering matrix $S$, the $K$-matrix is real, symmetric, and diagonal in $\ell$ and $n_{\mathrm{occ}}$. It is related to the $S$-matrix by  
\begin{align*}
K = (2 T^{-1} + i)^{-1}, \quad S = 1 + i T ,
\end{align*}
while for spinless particles $\tilde{K}^{-1}$ is defined as 
\begin{align}\label{e:ktl}
	\tilde{K}^{-1}_{\ell} (E_{\cm}) = \left(\frac{q_{\cm}}{m_\pi}\right)^{2\ell + 1} K^{-1}_{\ell} (E_{\cm}) = \left(\frac{q_{\cm}}{m_\pi}\right)^{2\ell + 1} \cot \, \delta_{\ell}(E_{\cm}) 
\end{align}
and is expected to be smooth near the elastic threshold. In this work the elements of 
$\tilde{K}$ are made dimensionless using $m_{\pi}$, which is a different 
convention for $\tilde{K}$ and $B$ compared to Eqs.~18 and 20 of Ref.~\cite{Morningstar:2017spu} which uses $2\pi/L$.  

When employing the determinant condition in Eq.~\ref{e:box} to the irreps listed in Tab.~\ref{t:irr}, partial wave mixing must be treated carefully. In order 
to proceed, we first neglect all partial waves with $\ell \ge 2$. The 
systematic error due to this truncation will be assessed shortly. After applying this restriction the $B$-matrices appearing in Eq.~\ref{e:box} are either 
one- or two-dimensional. For a one-dimensional $B$-matrix the determinant 
condition is of course trivial and yields a one-to-one relationship between 
a finite-volume energy $E_{\cm}$ and an amplitude point 
$\tilde{K}^{-1}_{\ell}(E_{\cm})$. This one-to-one relationship is typically 
exploited in calculations of elastic pion-pion scattering amplitudes. 

While there are a number of irreps listed in Tab.~\ref{t:irr} for which the $\ell=1$ partial wave can be isolated in this manner, is it only the $A_{1g}$ irrep 
at zero total momentum (denoted $A_{1g}(0)$) which provides unambiguous 
$s$-wave amplitude points. Therefore, we proceed with the determination of 
both amplitudes by simultaneously fitting the elastic energies in all irreps
according to the method of Ref.~\cite{Morningstar:2017spu}. For these global 
fits, a parametrization of the $s$- and $p$-waves are required which describes 
 $\tilde{K}^{-1}_{\ell}(E_{\cm})$ using a few fit parameters. These parameters are determined by 
minimizing a correlated $\chi^2$ which consists of residuals given by the 
determinants in Eq.~\ref{e:box}. Ref.~\cite{Morningstar:2017spu} also proposes 
another option for the residuals, namely 
\begin{align}\label{e:res}
	\Omega(\mu, A) = \frac{\mathrm{det}(A)}{\mathrm{det}
	[(\mu^2 + AA^{\dagger})^{1/2}]},
\end{align}
where $A = \tilde{K}^{-1} -B$ is the matrix appearing in the determinant, and 
$\mu$ is an arbitrary parameter chosen to suppress unimportant contributions to the determinant, which also improves the  
convergence of the minimization procedure.
The residuals in Eq.~\ref{e:res} are constructed to efficiently treat 
large-dimensional matrices, but we employ them here as consistency checks 
with the determinant-residual fits.  

Suitable parametrizations for these amplitudes are now discussed. Based on the 
expectation of a narrow $K^*(892)$ resonance, the $p$-wave amplitude is 
parametrized by a relativistic Breit-Wigner
\begin{align}\label{e:bw1}
	(\tilde{K}^{-1}_1)^{\textsc{bw}}(E_{\cm}) =
	\left(\frac{m^2_{K^*}}{m^2_{\pi}} - \frac{E^2_{\cm}}{m_{\pi}^2}\right)
	\frac{6\pi E_{\cm}}{g_{K^{*}\pi\pi}^2m_{\pi}}
\end{align}
resulting in fit parameters $m^2_{K^*}/m^2_{\pi}$ and $g_{K^{*}\pi\pi}^2$, both
of which are constrained to be non-negative. For the $s$-wave amplitude we 
employ a variety of parametrizations. Linear and quadratic parametrizations 
motivated by analyticity at threshold in $E_{\cm}$ and $s=E_{\cm}^2$ 
(respectively)
\begin{align}\label{e:lq}
	(\tilde{K}^{-1}_0)^\textsc{lin} (E_{\cm}) &= a_\textsc{lin} + b_\textsc{lin} E_{\cm} ,
	\\\nonumber 
	(\tilde{K}^{-1}_0)^\textsc{quad} (E_{\cm}) &= a_\textsc{quad} + b_\textsc{quad} E^2_{\cm}
\end{align}
each have two unconstrained fit parameters. We also include an $s$-wave parametrization including the first two terms in the effective range expansion 
\begin{align}\label{e:ere}
	(\tilde{K}^{-1}_0)^\textsc{ere} (q_{\cm}) &= \frac{-1}{m_{\pi}a_0}  +
	\frac{m_\pi r_0}{2} \frac{q^2_{\cm}}{m_{\pi}^2}
\end{align}
which depends on $q_{\cm}^2$ (rather than $E_{\cm}$) and contains two unconstrained fit parameters $m_{\pi} a_0$ and $m_{\pi}r_0$. 
In addition to these near-threshold 
parametrizations, we also explore an $\ell=0$ relativistic Breit-Wigner  
\begin{align}\label{e:bw0}
	(\tilde{K}^{-1}_0)^{\textsc{bw}}(E_{\cm}) =
	\left(\frac{m^2_{K_0^*}}{m^2_{\pi}} - \frac{E^2_{\cm}}{m_{\pi}^2}\right)
	\frac{6\pi m_{\pi} E_{\cm}}{g_{K_0^{*}\pi\pi}^2m^2_{K_0^*}}
\end{align}
with (non-negative) parameters $m^2_{K_0^*}/m^2_{\pi}$ and $g_{K_0^{*}\pi\pi}^2$.

We turn finally to assessment of the systematic error from the truncation to $\ell<2$. To this end, the determinant condition is simply enlarged to include a 
$d$-wave parametrized by the leading-order effective range expansion  
\begin{align}\label{e:dwv}
	(\tilde{K}^{-1}_2)^{\rm LO}(E_{\cm}) = -\frac{1}{m_{\pi}^5 a_2}
\end{align}
which contains a single unconstrained parameter $m_{\pi}^5 a_2$. It should be
stressed that if $\ell=2$ partial wave mixing is included, the only irreps 
which provide one-to-one determinations of the $\ell=0$ and $\ell=1$ 
amplitudes are the $A_{1g}(0)$ and $T_{1u}(0)$, respectively. 

\section{Results}\label{s:res}

The formalism discussed in Sec.~\ref{s:meth} for determining the finite-volume
energies and relating them to the infinite-volume elastic scattering 
amplitude is applied in this section. Results for the finite-volume energies, 
$B$-matrix elements (see Eq.~\ref{e:box}), and fit parameters for 
$\tilde{K}^{-1}_{\ell}(E_{\cm})$ are provided.

\subsection{Finite volume energies} 

Before discussing the results 
of the ratio fits which are used in the final amplitude analysis, 
exponential fits to $\hat{C}_n(t)$ are employed to investigate the overlaps 
of the finite-volume energies onto each of the interpolating 
operators. While these exponential fits are sometimes less precise and generally suffer from larger excited state contamination compared to ratio fits, they 
do not require knowledge of suitable nearby non-interacting states and are 
thereby used to verify ansatze for the ratio fits. 
   To this end, the GEVP eigenvectors from the operator bases listed in 
	 Tab.~\ref{t:ops} are used to form the overlaps 
	 \begin{align}\label{e:ovr}
		 Z_{in}(t) = \left| \frac{\sum_{j} C_{ij}(t) \, v_{nj} }{
			 \mathrm{e}^{-E_n t/2}\sqrt{\hat{C}_n(t)}}\right|^2  
	 \end{align}
	 where $E_n$ is the energy obtained from single-exponential fits and 
	 $v_{ni}$ the $i$th component of the $n$th GEVP eigenvector. The $Z_{in}(t)$
	 (apart from GEVP systematics) plateau to $Z_{in} = |\langle 0 | 
	 \hat{\mathcal{O}}_i | n \rangle |^2$. 

The finite-volume energies from these exponential fits, which 
are not those used in the final amplitude analysis, are displayed in 
Fig.~\ref{f:box} together with the overlaps of Eq.~\ref{e:ovr}. The overlaps 
of each interpolating operator onto a single finite-volume eigenstate 
are typically sharply peaked, indicating that each eigenstate has large 
overlap onto only one or two interpolating operators.  
\begin{figure}
	\centering
	\includegraphics{figures/box_plot.tikz}
	\caption{\label{f:box}All finite-volume two-hadron energies boosted to the 
	center-of-mass frame determined from single-exponential fits. Each irrep is 
	located 
	in one column, where the energies are shown in the upper panel as boxes with a 
	vertical dimension equal to the statistical error, the non-interacting two-hadron levels as solid horizontal lines, and the relevant thresholds as dashed gray lines. 
	The corresponding columns 
	in the lower panel indicate the overlaps (defined in Eq.~\ref{e:ovr}) of each
	interpolating operator onto the finite-volume Hamiltonian eigenstates. Ratio fits to those levels below $K\pi\pi$ threshold are used in the final analysis.} 
\end{figure}
Fig.~\ref{f:box} also demonstrates that the extraction of a few levels above 
 $K\pi\pi$ threshold is possible, however these levels do not have a 
straight-forward interpretation in terms of infinite-volume scattering 
amplitudes and are therefore not used in our final analysis. Although the first excited state in the $A_{1g}(0)$ irrep is just below the non-interacting $K\pi\pi$ energy, this threshold does not appear in that irrep so this level may 
be safely used in the analyis. 

For the final amplitude analysis, we instead employ the ratio fits. After 
reconstructing $E_{\cm}$ according to Eqs.~\ref{e:erf} and~\ref{e:kin}, it is used to calculate the 
$B$-matrix elements of Eq.~\ref{e:box}. Depending on the irrep in question, 
this matrix is either one- or two-dimensional if $\ell\ge 2$ contributions 
are ignored. 
\begin{table}
	\normalsize%

\begin{tabular}{c c c c c c c c}%
	\toprule
	$\phantom{^2}\dvec^2$&$\Lambda$&$E_{\rm cm}/m_\pi$&$\tilde{E}$&$B_{00}$&$B_{11}$&$\mathrm{Re}\, B_{01}$&$\mathrm{Im}\, B_{01}$\\%
\midrule%
 0&$A_{1g}$&$3.090(11)$&$0.9696(23)$&$3.05(29)$&---&---&---\\%
 &&$4.087(23)$&$1.966(18)$&$1.48(14)$&---&---&---\\%
\midrule%
 &$T_{1u}$&$3.787(30)$&$1.666(26)$&---&$0.094(72)$&---&---\\%
\midrule%
\midrule%
 1&$A_1$&$3.216(12)$&$1.0955(46)$&$1.97(26)$&$1.28(15)$&${-}1.57(19)$&$0.0$\\%
 &&$3.543(16)$&$1.4226(81)$&$0.73(19)$&$2.07(39)$&$2.36(25)$&$0.0$\\%
 &&$3.875(23)$&$1.754(18)$&${-}0.299(53)$&${-}1.438(65)$&${-}0.381(18)$&$0.0$\\%
\midrule%
 &$E$&$3.848(24)$&$1.727(20)$&---&$0.093(52)$&---&---\\%
\midrule%
\midrule%
 2&$A_1$&$3.346(13)$&$1.2253(72)$&$3.4(1.0)$&$3.6(1.1)$&$0.0$&$3.5(1.0)$\\%
 &&$3.728(22)$&$1.607(17)$&$1.59(41)$&$0.37(21)$&$0.0$&${-}1.70(19)$\\%
 &&$3.802(18)$&$1.682(10)$&$21(29)$&$4.4(3.9)$&$0.0$&${-}9(11)$\\%
 &&$3.935(23)$&$1.814(17)$&${-}3.04(82)$&${-}7.4(2.4)$&$0.0$&$3.5(1.3)$\\%
\midrule%
 &$B_1$&$3.814(27)$&$1.694(22)$&---&${-}0.107(40)$&---&---\\%
\midrule%
 &$B_2$&$3.676(20)$&$1.555(12)$&---&$2.21(28)$&---&---\\%
 &&$3.996(21)$&$1.876(15)$&---&${-}3.56(20)$&---&---\\%
\midrule%
\midrule%
 3&$A_1$&$3.436(15)$&$1.315(12)$&$2.6(1.0)$&$3.7(1.5)$&$2.20(86)$&$2.20(86)$\\%
 &&$3.806(37)$&$1.686(32)$&$1.06(35)$&${-}0.39(20)$&${-}0.699(13)$&${-}0.699(13)$\\%
 &&$3.925(32)$&$1.805(27)$&$4.3(2.5)$&$0.79(48)$&${-}0.66(20)$&${-}0.66(20)$\\%
\midrule%
 &$E$&$3.758(41)$&$1.638(34)$&---&$1.26(36)$&---&---\\%
 &&$4.056(28)$&$1.936(23)$&---&${-}2.9(3.0)$&---&---\\%
\midrule%
\midrule%
 4&$A_1$&$3.192(13)$&$1.0709(52)$&$2.68(67)$&$1.24(27)$&$1.65(42)$&$0.0$\\%
 &&$3.484(19)$&$1.363(15)$&$0.67(31)$&$1.76(54)$&${-}1.84(37)$&$0.0$\\%
 &&$3.721(55)$&$1.601(53)$&${-}0.77(32)$&${-}1.88(48)$&$0.99(26)$&$0.0$\\%
 \midrule
\end{tabular}%

	\caption{\label{t:res}Finite-volume two-hadron energies 
	in the center-of-mass frame (obtained from ratio fits) together with the 
	corresponding box matrix elements for $\ell<2$, which are defined in 
	Eq.~\ref{e:box}. A vanishing matrix element is denoted with a long dash, 
	while some off-diagonal elements are either exactly real or imaginary, 
	with the other component denoted by `$0.0$'. The information included here 
	is used to perform fits using various $K$-matrix parametrizations.}
\end{table}
The finite-volume energies obtained from the ratio fits and the $\ell=0,1$ 
$B$-matrix elements are displayed in Tab.~\ref{t:res}. This table contains all 
information (apart from an estimate of covariances, which may be provided on 
request) required to perform fits to 
determine $\tilde{K}_{\ell}(E_{\cm})$.

\subsection{$K$-matrix fits}

\begin{table}
\centering
\begin{tabular} {c c c c c c}
  \toprule
	Fit &	$s$-wave par. & $m_{K^*}/m_{\pi}$ & $g_{K^*K\pi}$ & $m_{\pi}a_0$ & $\chi^2/\mathrm{d.o.f.}$ \\
	\midrule
	(1a,1b) &	 \textsc{lin}   & $3.819(20)$ & $5.54(25)$ & $-0.333(31)$ & ($1.04,$\textendash) \\
	2 &	  \textsc{lin}   & $3.810(18)$ & $5.30(19)$ & $-0.349(25)$ & $1.49$ \\
	3 &	  \textsc{quad}   & $3.810(18)$ & $5.31(19)$ & $-0.350(25)$ & $1.47$ \\
	4 &	  \textsc{ere}   & $3.809(17)$ & $5.31(20)$ & $-0.351(24)$ & $1.47$ \\
	5 &	  \textsc{bw}   & $3.808(18)$ & $5.33(20)$ & $-0.353(25)$ & $1.42$ \\
	6 &	  \textsc{bw}   & $3.810(17)$ & $5.33(20)$ & $-0.354(25)$ & $1.50$ \\
	\bottomrule
\end{tabular}
	\caption{\label{t:fit}Results for the $K^*(892)$ resonance parameters and the 
	$s$-wave scattering length $m_{\pi}a_0$ from all fits to the amplitudes.
	 For each fit, the $p$-wave amplitude is described using the Breit-Wigner of 
	 Eq.~\ref{e:bw1}.
	The first row contains results from independent fits to $\ell = 0,1$ separately, denoted (1a,1b), using only irreps without $\ell=0,1$ mixing. This yields a meaningless $\chi^2/\mathrm{d.o.f.}$ for the $s$-wave since there are only two elastic $A_{1g}(0)$ 
	levels. Fit 6 includes $d$-wave contributions as discussed in the text.}  
\end{table}
Using the ratio fits and box matrix elements collated in Tab.~\ref{t:res}, we 
turn now to fitting $\tilde{K}_{\ell}(E_{\rm cm})$ according to the method
outlined in Sec.~\ref{s:amp}. As a basis for comparison, we first employ only 
irreps in which there is 
no $\ell=0,1$ partial wave mixing and perform independent fits to the $s$- and 
$p$-waves separately using the Breit-Wigner form of Eq.~\ref{e:bw1} for 
$\ell=1$ and the linear form of Eq.~\ref{e:lq} for $\ell=0$. These fits are 
denoted $(1a,1b)$ in Tab.~\ref{t:fit}. Since there are only two $A_{1g}(0)$ levels in 
this two-parameter linear $s$-wave fit, the $\chi^2/\mathrm{d.o.f.}$ is 
meaningless.  

Next we consider simultaneous fits to both $\ell=0,1$ partial waves. As in fit 1a, the $p$-wave is always described by Eq.~\ref{e:bw1} in these fits, which 
are also listed in Tab.~\ref{t:fit}. Fit 2 employs the linear $s$-wave form, 
fit 3 the quadratic one from Eq.~\ref{e:lq}, fit 4 the NLO effective range expansion of Eq.~\ref{e:ere} (yielding $m_{\pi}r_0 = -1.74(31)$), and fit 5 the 
$s$-wave 
Breit-Wigner of Eq.~\ref{e:bw0}. Fit 6 also employs the $s$-wave Breit-Wigner 
but enlarges the $\tilde{K}$- and $B$-matrices to include $d$-wave contributions
according to Eq.~\ref{e:dwv}. Together 
with the parameters listed in Tab.~\ref{t:fit}, fit 6 constrains the $d$-wave contribution to be $m_{\pi}^5a_2 = -0.0013(68)$. 

As is evident from Tab.~\ref{t:fit}, the $K^*(892)$ resonance parameters are 
insensitive to the $s$-wave parametrization and the inclusion of 
$d$-wave contributions. Similarly, if each of the $s$-wave parametrizations are 
used to interpolate to $K\pi$ threshold and determine the scattering length 
$m_{\pi}a_0$, the resulting values also do not vary significantly with 
different parametrizations or the inclusion of the $d$-wave.  

\begin{figure}[!ht]
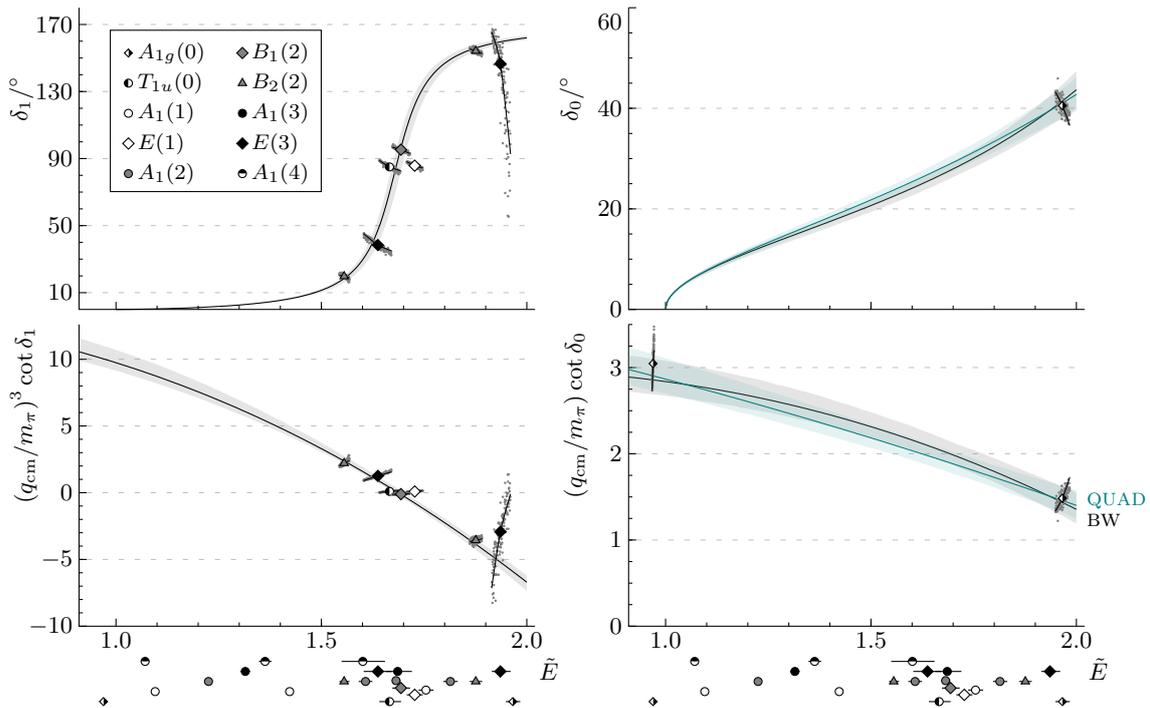

  	\ifdefined\BUILDFIGS
	\includegraphics{figures/ell1plot.tikz}
	\hspace{-1.2em}
	\includegraphics{figures/ell0plot.tikz}
	\fi
	\caption{\label{f:cot}$K$-matrix fits to the $s$- and $p$-wave amplitudes.
	Together with the fits, which are explained in the text, we show amplitude 
	points (neglecting $d$-wave contributions)
	from irreps which do not mix these two partial waves. All energies involved
	in the fit are indicated below the plots where they are offset vertically for clarity.}
\end{figure}

The amplitudes from fit 3 are shown in Fig.~\ref{f:cot}, together with
the Breit-Wigner $s$-wave amplitude from fit 5, illustrating that
different parametrizations 
for the $s$-wave produce a similar energy dependence in the elastic region. In addition to the 
fits, points from irreps without $\ell=0,1$ partial wave mixing are shown
and seen to be consistent. 

We now briefly discuss the $s$-wave amplitude in the context of the 
$K_{0}^{*}(800)$. Based on the LO effective range expansion, a negative $m_{\pi}a_0$ suggests a virtual bound state. However, $q_{\cm}\cot \delta_0$ has 
a significant slope, as is evident in Fig.~\ref{f:cot}. 
The NLO effective range parameters of fit 4 are used to construct the ratio 
$1-2r_0/a_0 = -8.9(2.4)$, which must be positive in the presence of a (real or 
virtual) bound state. A near-threshold bound state is therefore disfavored at 
the $3-4\sigma$ level.    

The existence of a resonance pole above threshold on the lower half of the 
second (unphysical) Riemann sheet requires a careful analytic continuation, 
and most likely a better energy resolution than we have achieved here. 
Nonetheless, qualitative information about a possible $s$-wave pole may 
obtained by determining the zeros of $q_{\cm}\cot \delta_0 -iq_{\cm}$. This is 
easily done using the NLO effective range parametrization of fit 4 and solving 
the resultant quadratic polynomial, yielding
$m_{R}/m_{\pi} = 4.66(13) -0.87(18)i$ which is consistent with the Breit-Wigner 
mass and width from fit 5, which gives $m_{K_{0}^{*}}/m_{\pi} = 4.59(11)$ and 
$g_{K_0^{*}K\pi} = 3.35(17)$. It should be noted that in 
addition to the $K_{0}^{*}(800)$, the $s$-wave amplitude may also be 
influenced by the $K_0^{*}(1430)$ resonance. Overall, without 
a full analytic continuation
we can only infer qualitative information about a possible 
$s$-wave resonance pole from the elastic amplitude calculated here.  

\section{Conclusions}\label{s:concl} 

In this work $22$ finite-volume $K\pi$ energies calculated from an 
$\nf=2+1$ lattice QCD simulation are employed to determine the $I=1/2$, 
$S=1$ elastic $s$- and $p$-wave $K\pi$ scattering amplitudes. Due to the 
scattering of non-identical particles, both of these partial waves contribute to finite-volume
 two-hadron energy shifts in some irreps. We treat this partial 
wave mixing by fitting both $\ell=0,1$ contributions simultaneously, 
while the $K^{*}(892)$ resonance parameters and the $s$-wave scattering length 
are insensitive to the parametrization chosen for the $s$-wave and 
the inclusion of $d$-wave contributions.  

\begin{figure}
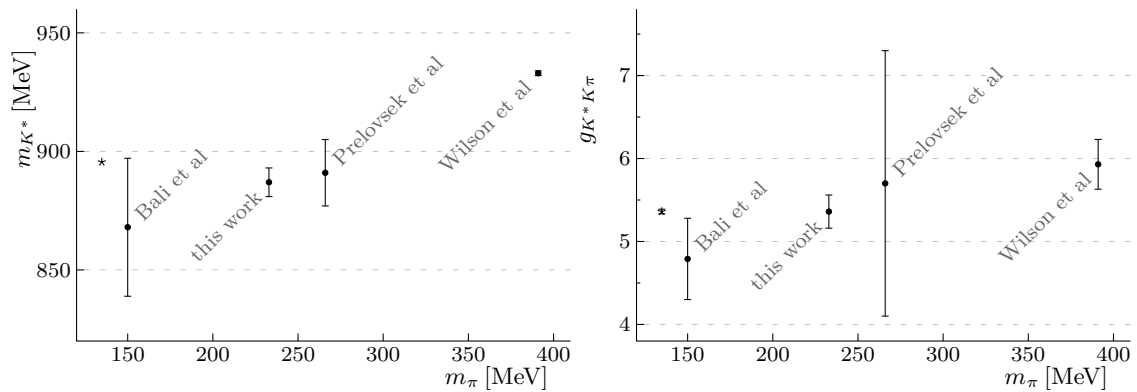

  \centering
  \includegraphics{figures/summPlot_mr_mpi.tikz}
  \hspace{-0.5em}%
  \includegraphics{figures/summPlot_g_mpi.tikz}
	\caption{\label{f:kst} Summary of lattice QCD calculations of $K^{*}(892)$ 
	resonance parameters, together with phenomenological values 
	(shown as asterisks) from
	Ref.~\cite{Patrignani:2016xqp} where the neutral values for the mass and width are taken. This choice gives consistent values to 
	Ref.~\cite{Bali:2015gji}, while hadro-produced $K^{*}(892)$ parameters result in a coupling which is about $5\%$ larger. The statistical and systematic
	errors from Ref.~\cite{Wilson:2015dqa} are added in quadrature.}
\end{figure}

For our values for these quantities we take fit 5 from Tab.~\ref{t:fit}
\begin{align} 
	\frac{m_{K*}}{m_{\pi}} = 3.808(18), \quad g_{K^{*}K\pi} = 5.33(20), 
	\quad m_{\pi}a_0 = -0.353(25), 
\end{align}
where the errors are statistical only. These values may be compared with 
existing $K^{*}(892)$ resonance 
calculations~\cite{Prelovsek:2013ela,Wilson:2015dqa,Bali:2015gji}. 
Ref.~\cite{Prelovsek:2013ela} employs a single $\nf=2$ ensemble with 
similar (albeit somewhat 
heavier) pion mass of $m_{\pi} = 266\mathrm{MeV}$ in a smaller spatial volume with 
$L=2\mathrm{fm}$, resulting in four elastic levels from irreps which do not mix 
with $\ell=0$. Ref.~\cite{Wilson:2015dqa} employs a larger pion mass of 
$m_{\pi}=390\mathrm{MeV}$ and three $\nf=2+1$ ensembles with spatial extents  
in the range $L=2-3\mathrm{fm}$. At these quark masses the $\eta K$ threshold 
is below $\pi\pi K$, resulting in about $50$ two-hadron levels, 
and a full coupled-channel analysis is performed. Although  
the $K^*(892)$ is stable but close to threshold at this heavy light quark mass, 
$q_{\cm}^3 \cot \delta_1$ is nevertheless well described by a Breit-Wigner 
shape, while analytic continuation of the $s$-wave amplitude 
parametrizations suggests a virtual bound state corresponding to the $K_{0}^{*}(800)$.
Finally, Ref.~\cite{Bali:2015gji} employs two $\nf=2$ ensembles with 
different pion masses ($m_{\pi}=150,\,160\mathrm{MeV}$) near the physical point  and 
spatial extents $L=3.5-4.6\mathrm{fm}$ in a single amplitude fit. Due to these
light quark masses, almost no levels are in the elastic region. A summary 
of these existing results on  
$K^{*}(892)$ resonance parameters is shown in Fig.~\ref{f:kst} together with 
our results. When considering this comparison, one must keep in mind
the different 
scale setting and strange quark mass tuning procedures. 

Despite the lack of theoretical control over levels above $K\pi\pi$ threshold,
all three of Refs.~\cite{Prelovsek:2013ela,Wilson:2015dqa,Bali:2015gji} employ 
a number of such levels in their amplitude analysis. While one may be tempted 
to argue that some inelastic levels in Fig.~\ref{f:box} have strong overlap
with two hadron interpolators and therefore are well-described by the 
two-hadron quantization condition of Eq.~\ref{e:box}, 
Ref.~\cite{Giudice:2012tg} provides an example (shown there in Fig.~16) where 
this is not the case. In principle, the formalism developed in 
Ref.~\cite{Briceno:2017tce} could be adapted to non-identical scalar particles
to investigate the magnitude of this systematic error.

For the $s$-wave amplitude, calculations of $m_{\pi}a_0$ are considerably 
more mature than those of the $K^{*}(892)$ resonance 
parameters~\cite{Fu:2011wc,Janowski:2014uda,Lang:2012sv,Beane:2006gj,Sasaki:2013vxa}. The state-of-the-art for these calculations involves an extrapolation
to the continuum and physical quark masses, so comparison with our value is not
appropriate. Nonetheless, our value of $m_{\pi}a_0 = -0.353(25)$ is consistent with expectations from chiral effective theory, shown in Fig.~10 of 
Ref.~\cite{Fu:2011wc}. 

As mentioned above, on this single ensemble we are unable to estimate the magnitude of lattice 
spacing effects and exponentially suppressed finite-volume corrections to 
Eq.~\ref{e:box}, nor are we able to extrapolate the light quark masses to 
their physical values. This will require a large set of ensembles, such 
as the CLS ensembles currently employed for an ongoing  
calculation of the $I=1$ elastic $\pi\pi$ amplitude which aims to assess 
these effects~\cite{Bulava:2017stw}. 

Nonetheless, the results reported here are a valuable proof-of-principle and 
demonstrate the statistical precision which may be attained in such future 
calculations, although levels near the $K^{*}(892)$ may exhibit exponential signal-to-noise related degradation in 
precision as the physical light quark mass is approached. Perhaps a more relevant issue is the decrease of the 
elastic energy region as the $K\pi\pi$ threshold is lowered to its physical 
value. To our knowledge, the three-body formalism of 
Ref.~\cite{Briceno:2017tce} has not yet been applied to numerical lattice data. 

Apart from approaching the physical point directly in the lattice simulations, 
information about physical scattering amplitudes may be inferred from lattice
data at heavier quark masses which are however still in the range of 
applicability of chiral effective theories. This novel interplay between 
 effective field theories and resonant lattice scattering 
data is currently underway~\cite{Bavontaweepanya:2018yds,Hu:2017wli,
Doring:2016bdr,
Bolton:2015psa,Liu:2016wxq,Liu:2016uzk,MartinezTorres:2017bdo,Guo:2016zep,
Guo:2018kno}. 

The technology underlying the simultaneous fits performed here to different $K$-matrix elements is 
similar to the treatment required for coupled channel problems, on which 
first calculations have appeared for the $K\pi-K\eta$, $\eta\pi-\bar{K}K$, and 
$\pi\pi-\eta\eta-\bar{K}K$ 
systems at a heavier pion mass and smaller physical volume than
this work~\cite{Wilson:2015dqa}. 
The methods used here may also be taken over to meson-baryon systems, where the
non-zero intrinsic spin provides an additional complication. Nonetheless, 
first progress on resonant nucleon-pion scattering has been reported recently 
in Ref.~\cite{Andersen:2017una}.

\vspace{4mm}
\noindent
{\bf Acknowledgements}: 
We gratefully acknowledge conversations with D. Mohler and M.F.M. Lutz. This work was supported by the U.S.~National Science Foundation 
under award PHY-1613449.  Computing resources were provided by
the Extreme Science and Engineering Discovery Environment (XSEDE)
under grant number TG-MCA07S017.  XSEDE is supported by National 
Science Foundation grant number ACI-1548562. The USQCD QDP++ 
library~\cite{Edwards:2004sx} and the Improved
BiCGStab solver in Chroma were used in developing
the software for early stages of the calculations reported here.

\appendix
\section{$\tmin$-plots for moving kaons}\label{a:k} 

\begin{figure}
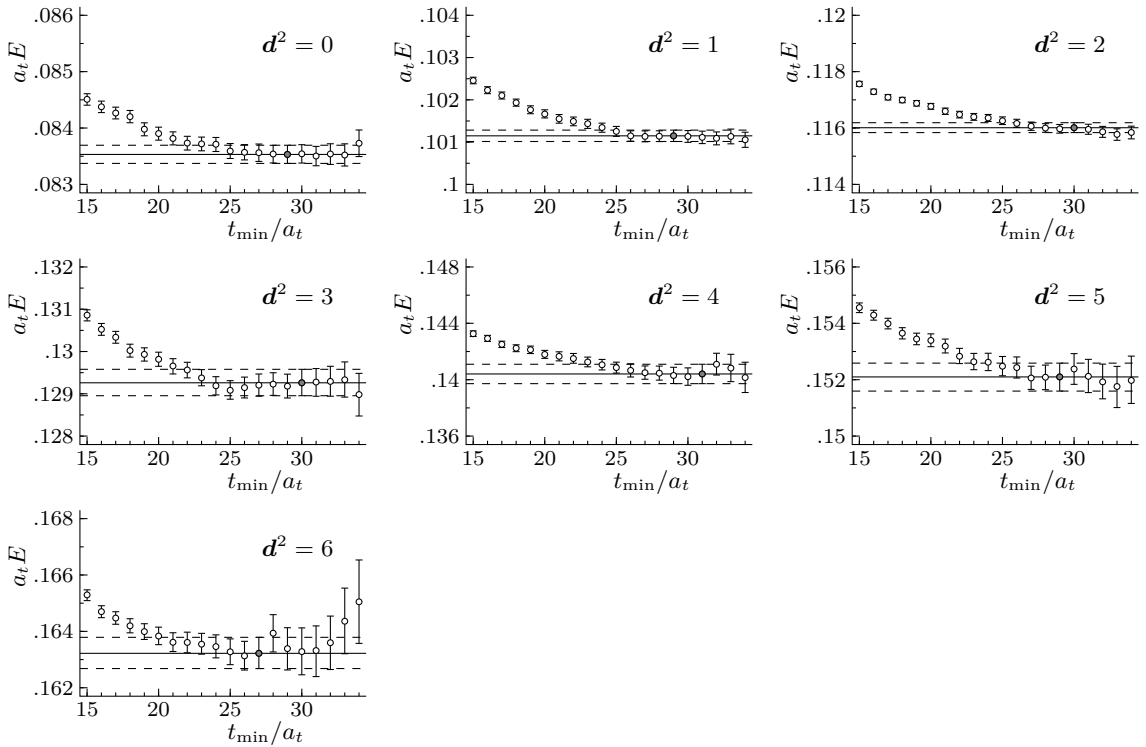

  \ifdefined\BUILDFIGS
  \includegraphics{figures/kaon_pSq0_tmin.tikz}
  \hspace{-0.25cm}
  \includegraphics{figures/kaon_pSq1_tmin.tikz}
  \hspace{-0.25cm}
  \includegraphics{figures/kaon_pSq2_tmin.tikz}
  \includegraphics{figures/kaon_pSq3_tmin.tikz}
  \hspace{-0.25cm}
  \includegraphics{figures/kaon_pSq4_tmin.tikz}
  \hspace{-0.25cm}
  \includegraphics{figures/kaon_pSq5_tmin.tikz}
  \includegraphics{figures/kaon_pSq6_tmin.tikz}
  \fi
	\caption{\label{f:xik} Plots of the $\tmin$ dependence of the fitted energies 
	for all moving kaons used to determine $\xi_K$ quoted in Tab.~\ref{t:ens} and Fig.~\ref{f:keta}. For all fits, the maximum time separation is 
	$\tmax = 38a_t$ while the chosen fit is indicated with a filled symbol.}  
\end{figure}

This appendix contains $\tmin$-plots for single exponential fits to each 
of the moving kaon correlation functions used in determination of the kaon anisotropy $\xi_K$ 
discussed in Sec.~\ref{s:ens}, which are shown in Fig.~\ref{f:xik}. As shown in
Fig.~\ref{f:keta}, this determination of $\xi_K$ is consistent with $\xi_{\pi}$ 
determined previously in Ref.~\cite{Bulava:2016mks}. 

\section{Operator bases for all irreps}\label{a:ops}

We detail here the basis of interpolating operators used in solving the GEVP of Eq.~\ref{e:gevp} in every irrep. Each operator is constructed to transform 
irreducibly according to a particular irrep, as detailed in 
Ref.~\cite{Morningstar:2013bda}. While various classes of covariantly 
displaced operators 
 are considered in Ref.~\cite{Morningstar:2013bda}, only 
single-site operators are used here. For each spatial displacement type, 
a number of linearly independent operators were determined in 
Ref.~\cite{Morningstar:2013bda}, each of which is identified by a spatial 
identification number placed after the spatial displacement type, such as 
$\mathrm{SS}0$ for the zeroth single-site operator in a particular irrep.

When forming $K\pi$ correlation functions, there is some freedom in choosing 
the interpolating operators for the constituent pion and kaon. Here we always 
choose the $\mathrm{SS}0$ operator for all pion and kaon interpolators inside 
our $K\pi$ operators, except for those with a single unit of momentum where we 
use the $\mathrm{SS}1$ operators. These compound operators are therefore 
denoted $K(\boldsymbol{d}^2_{K})\pi(\boldsymbol{d}^2_{\pi})$, where the 
displacement type and spatial identification number are implied and the 
integers in 
parenthesis are momenta given in units of $2\pi/L$. Operator identifiers 
are indicated explicitly for our single-hadron interpolators in 
Tab.~\ref{t:ops}. 
\begin{table}
		\centering
\begin{tabular}{c c c}
  \toprule
  $\dvec^2$ & \textbf{$\Lambda$} & operators \\
  \midrule
	$0$ &$A_{1g}$& $K(0)\pi(0),\, K(1)\pi(1),\, K(2)\pi(2),\, K(3)\pi(3),\, K(0)_{\mathrm{SS}0}$  \\
	&$T_{1u}$&  $K(0)_{\mathrm{SS}1},\, K(1)\pi(1),\, K(2)\pi(2)$ \\
\midrule%
	$1$ &$A_1$&  $K(1)\pi(0),\, K(0)\pi(1),\, K(1)_{\mathrm{SS}2},\, K(1)_{\mathrm{SS}0},\, K(2)\pi(1),\, K(1)\pi(2)$ \\
	&$E$& $K(1)_{\mathrm{SS}2},\, K(1)\pi(2),\, K(2)\pi(1)$ \\
\midrule%
	$2$ &$A_1$& $K(2)\pi(0),\, K(1)\pi(1),\, K(2)_{\mathrm{SS}3},\, K(0)\pi(2),\, K(3)\pi(1),\, K(2)\pi(2),\, K(1)\pi(3)$ \\
	&$B_1$& $K(2)_{\mathrm{SS}1},\, K(3)\pi(1),\, K(1)\pi(3),\, K(2)\pi(2)$ \\
	&$B_2$& $K(2)_{\mathrm{SS}3},\, K(1)\pi(1),\, K(2)\pi(2),\,K(2)_{\mathrm{SS}0}$ \\
\midrule%
	$3$ &$A_1$& $K(3)\pi(0),\, K(3)_{\mathrm{SS}3},\, K(2)\pi(1),\, K(0)\pi(3),\, K(1)\pi(2)$ \\
	&$E$& $K(3)_{\mathrm{SS}1},\, K(2)\pi(1),\, K(1)\pi(2)$ \\
\midrule%
	$4$	&$A_1$& $K(1)\pi(1),\, K(4)\pi(0),\, K(4)_{\mathrm{SS}0},\, K(4)_{\mathrm{SS}2},\, K(0)\pi(4),\, K(2)\pi(2)$ \\
\bottomrule
\end{tabular}

	\caption{\label{t:ops}Operator bases included in the GEVP for each two-hadron
	irrep. Each single-hadron operator is specified by a displacement type and
	a spatial identification number while the `K' refers only to the flavor 
	structure. The operators used for
	kaons and pions appearing in two-hadron operators are discussed in the text.
	The momentum of each operator (in units of $2\pi/L$) squared is shown in
	parenthesis.}
\end{table}

Coefficient files defining these operators are available upon request, and 
the list of operators used in each irrep is given in Tab.~\ref{t:ops}. Each 
list contains $n_{\rm op}$ operators, while a basis of size $n_{\rm op}-1$ 
 is obtained by removing the last operator in the list. These two different 
 bases are used to monitor the stability of the energies, as shown in 
 Fig.~\ref{f:fit2}.

\section{$\tmin$-plots for all two-hadron levels}

\begin{figure}
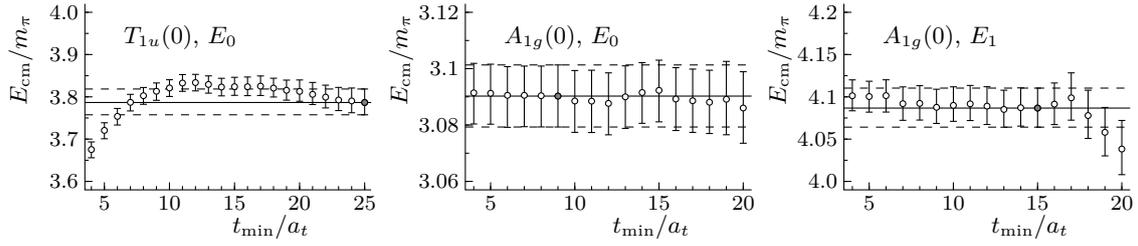

  \ifdefined\BUILDFIGS
  \includegraphics{figures/pSq0-T1u0_tmin.tikz}
  \hspace{-0.25cm}
  \includegraphics{figures/pSq0-A1g0_tmin.tikz}
  \hspace{-0.25cm}
  \includegraphics{figures/pSq0-A1g1_tmin.tikz}
  \fi
  \caption{\label{f:tm0}$\tmin$-plots of center-of-mass energy $E_\mathrm{cm}$ 
	for $\boldsymbol{d}^2 = 0$. The fit value for the chosen $\tmin$ is indicated by the error band and the filled circle.}
\end{figure}

Here we show $\tmin$-plots from ratio fits to all two-hadron levels in the 
amplitude analysis, which employ the GEVP's specified in App.~\ref{a:ops}.
For the $\boldsymbol{d}^2=0$ irreps, $\tmax = 26a_t$ is employed, while all 
other fits use $\tmax = 35a_t$. Total momentum zero levels are shown in 
Fig.~\ref{f:tm0} and those with $\boldsymbol{d}^2=1,2,3,4$ in 
Figs.~\ref{f:tm1},\ref{f:tm2},\ref{f:tm3},\ref{f:tm4}, respectively.  
\begin{figure}
  \ifdefined\BUILDFIGS
  \includegraphics{figures/pSq1-A10_tmin.tikz}
  \hspace{-0.25cm}
  \includegraphics{figures/pSq1-A11_tmin.tikz}
  \hspace{-0.25cm}
  \includegraphics{figures/pSq1-A12_tmin.tikz}
  \includegraphics{figures/pSq1-E0_tmin.tikz}
  \fi
	\caption{\label{f:tm1}Same as Fig.~\ref{f:tm0}, but for all levels with 
	$\boldsymbol{d}^2 = 1$.}
\end{figure}
\begin{figure}
  \ifdefined\BUILDFIGS
  \includegraphics{figures/pSq2-A10_tmin.tikz}
  \hspace{-0.25cm}
  \includegraphics{figures/pSq2-A11_tmin.tikz}
  \hspace{-0.25cm}
  \includegraphics{figures/pSq2-A12_tmin.tikz}
  \includegraphics{figures/pSq2-A13_tmin.tikz}
  \hspace{-0.25cm}
  \includegraphics{figures/pSq2-B10_tmin.tikz}
  \hspace{-0.25cm}
  \includegraphics{figures/pSq2-B20_tmin.tikz}
  \includegraphics{figures/pSq2-B21_tmin.tikz}
  \fi
	\caption{\label{f:tm2}Same as Fig.~\ref{f:tm0}, but for 
	$\boldsymbol{d}^2 = 2$.}
\end{figure}
\begin{figure}
  \ifdefined\BUILDFIGS
  \includegraphics{figures/pSq3-A10_tmin.tikz}
  \hspace{-0.25cm}
  \includegraphics{figures/pSq3-A11_tmin.tikz}
  \hspace{-0.25cm}
  \includegraphics{figures/pSq3-A12_tmin.tikz}
  \includegraphics{figures/pSq3-E0_tmin.tikz}
  \hspace{-0.25cm}
  \includegraphics{figures/pSq3-E1_tmin.tikz}
  \fi
	\caption{\label{f:tm3}Same as Fig.~\ref{f:tm0}, but for 
	$\boldsymbol{d}^2 = 3$.}
\end{figure}
\begin{figure}
  \ifdefined\BUILDFIGS
  \includegraphics{figures/pSq4-A10_tmin.tikz}
  \hspace{-0.25cm}
  \includegraphics{figures/pSq4-A11_tmin.tikz}
  \hspace{-0.25cm}
  \includegraphics{figures/pSq4-A12_tmin.tikz}
  \fi
	\caption{\label{f:tm4}Same as Fig.~\ref{f:tm0}, but for 
	$\boldsymbol{d}^2 = 4$.}
\end{figure}


\end{document}